\documentclass[aps,pre,floatfix,nofootinbib]{revtex4}
\usepackage{epsfig,ulem,color,subfigure}
\usepackage{amsmath,amsthm,amssymb}
\newcommand{\be}{\begin{equation}}
\newcommand{\ee}{\end{equation}}

\begin{document}
\title{Modeling the formation of {\it in vitro} filopodia}
\date{\today}
\author{K.-C. Lee$^1$, A. Gopinathan$^2$, and J. M. Schwarz$^{3,}$\footnote[4]{E-mail address of corresponding author: jschwarz@physics.syr.edu}}
\affiliation{$^1$Department of Mathematics and Department of Neurobiology, Physiology and Behavior, University of California-Davis, Davis, CA 95616\\ 
$^2$School of Natural Sciences, University of California-Merced, Merced, CA 95348\\
$^3$Physics Department, Syracuse University, Syracuse, NY 13244}
\begin{abstract}  
Filopodia are bundles of actin
filaments that extend out ahead of the leading edge of a crawling cell to probe its
upcoming environment. {\it In vitro} experiments [D. Vignjevic {\it et al.}, J. Cell Biol. {\bf 160}, 951 (2003)] have determined the
minimal ingredients required for the formation of filopodia from the dendritic-like morphology of the leading edge. We model these experiments using kinetic aggregation equations for the density of growing bundle tips. In mean field, we determine the bundle size distribution to be broad for bundle sizes smaller than a characteristic bundle size above which the distribution decays exponentially.  Two-dimensional simulations incorporating both bundling and cross-linking measure a bundle size distribution that agrees qualitatively with mean field. The simulations also demonstrate a nonmonotonicity in the radial extent of the dendritic region 
as a function of capping protein concentration, as was observed in experiments, due to the interplay between percolation and the ratcheting of growing filaments off a spherical obstacle. 
\end{abstract}
\maketitle  
\section{Introduction}
Just as our skeletons give shape to our bodies, the cytoskeleton gives shape 
to the cell.  However, unlike our skeletons, the ``bones'', or filaments, of the cytoskeleton are dynamic---lengthening, shortening, branching, etc.--- so that the cell can extend itself to crawl 
along surfaces to search for food. The physical process of cell crawling consists of (1) formation of a broad, thin protrusion of actin filaments along part of the cell periphery (a lamellipod), (2) the assembly of adhesion sites adhering the protrusion to the surface and disassembly of adhesion sites in the ``rear'' of the cell, and (3) retraction of the rear to catch up with the ``front'' part where the new growth has occurred. 

To date there are a number of quantitative models addressing each aspect of cell motility~\cite{review1,review2,review3}.  For example, numerous observations of lamellipodia have led researchers to conjecture that lamellipodia are initiated by the branching agent Arp2/3, which nucleates new filaments from prexisting ones at an angle of about 70 degrees from the growing end of the polar actin filaments~\cite{mullins,pollard,borisy}.  The resulting structure forms a directed, branched/dendritic-like network of filaments at the leading edge.  Quantitative models implementing this mechanism of formation are able to account for observed cell speeds and various observed morphological properties~\cite{mogilner2,schaus,carlsson0,dawes,gopinathan}.

Here we quantitatively investigate a particular aspect of this emergent protrusion, namely the formation of bundles of actin filaments that extend out ahead of lamellipodia~\cite{mallavarapu}. These bundles are known as filopodia.  It has been proposed that filopodia serve as scouts for the cell~\cite{alberts}. Given this important role for filopodia in cell crawling, we ask the following questions: How do filopodia form?  More specifically, how do they emerge from 
a branched structure?  What is the typical number of actin filaments in a filopod?  To build a more comprehensive framework of filopodia formation, how do oriented bundles of actin filaments emerge in the absence of lamellipodia? 

To begin to address the formation of filopodia, {\it in vitro} experiments conducted by Vignjevic and collaborators~\cite{vignjevic}, along with experiments followed up by Haviv and collaborators~\cite{haviv} and Ideses and collaborators~\cite{ideses}, have determined the minimal ingredients required for filopodia formation. In the initial experiment~\cite{vignjevic}, a 
bead coated with Arp2/3 activator was inserted 
into a solution containing actin monomers, linker/bundling proteins, capping protein, and Arp2/3. After some minutes, for a large enough concentration of Arp2/3, actin filaments emerge and form a mesh-like phase close to the bead.  Under certain conditions, bundles of parallel filaments form further away from the bead---resulting in a star-like formation.  The more recent {\it in vitro} experiments by Haviv and collaborators~\cite{haviv} are conducted without a bead coated with Arp2/3 activator, but with activator in solution. They find similar star formation under certain conditions.  

Svitkina and collaborators~\cite{svitkina2} proposed two mechanisms for filopodia formation given the findings of the initial experiment.  The first is that filopodia are the result of cone-like structures, termed lambda precursors, consisting of filaments converging as they emerge from the dendritic network and continue on to grow and bundle.  Similar structures have also been observed in {\it in vivo} B16F1 melanoma cells~\cite{mejillano}.  Svitkina and collaborators~\cite{svitkina2} have dubbed this mechanism of formation, the convergent elongation model. A second mechanism consists of longer bundles coalescing laterally via zippering to form even larger bundles. This latter mechanism does not preclude the former, however, the latter is probably less likely given the larger-scale rearrangements required to bring two pre-formed bundles, each with their respective base structure, together. 
 
The notion of filaments elongating and bundling is a rather natural one beggingg for quantitative modelling.  We use the powerful framework of kinetic aggregation, initiated by the 1916 work of Smoluchowski studying the size distribution of ``sticky'' Brownian particles~\cite{smoluchowski}, to build a model for filopodia formation~\cite{leyvraz}.  We construct an equation of motion for filament tips that grow and bundle upon colliding as dictated by simple geometry with some caveats regarding the details of the bundling. We use the equation of motion for filament tips to compute the distribution of bundle sizes, a measureable quantity.  No energetics is incorporated into the model explicitly, an assumption that is supported by work by Yang and collaborators~\cite{yang}, where careful simulations of two constrained-at-one-end semiflexible polymers fluctuating and coalescing show that energetics denominates beyond a time scale of 10 seconds.  In the star system, filaments lengthen and nucleate on a much faster time scale and so we expect the star formation to be kinetically driven.

There have been several quantitative kinetic modeling papers addressing the formation of filopodia in the context of these experiments. For example, Mogilner and Rubinstein~\cite{mogilner} investigate the dynamics of lambda precursors colliding and maturing into full-blown filopodia. They find that density of filopodia at the leading edge is proportional to the square root of the initiation rate of lambda precursors. More recent work by Kraikivski and collaborators~\cite{kraikivski} investigates the tip-tip bundling events versus tip-side bundling events to find that, contrary to expectation, both types of events produce a comparable number of collisions.  However, the number of bundling events scales differently with the average bundle length for each respective event, as expected. 
 
The previous two theoretical works do not address a bundle size distribution. There exists a bundle size distribution calculation in a work, inspired by the initial experiments, treating the growing filaments as Potts-like spins on a lattice~\cite{stewman}. However, the kinetic Monte Carlo analysis yields a bundle size distribution with a prominent peak near a bundle size of three filaments (spins). This number is somewhat small compared with the qualitative estimates from the experiments. In the absence of Arp2/3, quantitative {\it in vitro} measurements of the bundle size distribution show that the average bundle size ranges from 10-30 depending on the ratio of actin to cross-linker concentration~\cite{haviv2}. Ref.~\cite{haviv2} also proposes a model for the bundle size distribution where there is a rate for filaments leaving the bundle as well as contributing to the bundle.  Since the filaments are entangled, extended objects, the rate for leaving a bundle seems somewhat small, however.

In this work, we calculate mean field and two-dimensional bundle size distributions, as opposed to the total number of bundling events. In mean field, we find that tip-tip bundling alone is not sufficient to generate substantive bundle sizes, i.e. an average bundle size of at least 10. In two-dimensions, we find a threshold concentration of filaments, unbranched or branched, for substantive bundling. In mean field and in two-dimensions we find that the bundle size distribution is broad for bundle sizes below some characteristic bundle size and exponentially decays for bundle sizes larger than the characteristic bundle size. 

We also probe the change in morphology from the dendritic ``phase'' to bundled ``phase'' and find that to observe a highly oriented/bundled phase (1) Arp2/3 needs to be excluded from the periphery of the system and (2) the bundling proteins must reorient the filaments/bundles.  We then investigate the effects of capping protein on the star formation and find a nonomonotonic dependence on the radial distribution of filament tips as a function capping protein in the presence of the bead. The bead is an obstacle which filaments growing into the bead can ratchet off of.  The higher the concentration of capping protein, the less likely a percolating structure will form across the system such that filaments, and their cross-linked and bundled neighbors, can ratchet off the bead and expand the dendritic region. 

The organization of the paper is as follows.  Section II constructs a general model based on the kinetic aggregation of bundle tips to calculate the bundle size distribution.  Section III addresses the model within mean field.  Section IV introduces and tests a kinetic simulation to go beyond mean field. Section V discusses the implications of our results. 
 
\section{Model}
Given the {\it in vitro} experiments, let us assume that there are 
two processes in the lamellipodia-filopodia transition: (1) branching by Arp2/3 and (2) bundling by cross-linking proteins.  Branching allows for the nucleation of filaments, while bundling coalesces smaller bundles into larger ones. We will also investigate the effects of capping to more closely model the initial {\it in vitro} filopodia experiment where the concentration of capping protein was varied. Capping results in ``death'' of a growing filament tip. 
These three processes are represented diagrammatically in Figure 1. In the 
branching process, a new filament emerges from a prexisting one. For the 
bundling/coalescing process, we assume that when two growing tips intersect 
with an angle less than $\theta_c$, the two bundles realign and 
bundle with probability unity, otherwise, no bundle forms~\cite{kierfeld,yu,yang}.  In principle, 
for intersection angles larger than $\theta_c$, a cross-link can still form without bundle reorientation.  This large-angle cross-link will be addressed in Section IV.  
There are other ways to bundle. In particular, the growing end of one bundle 
could collide into the side of another bundle.   We call this tip-side bundling. See Fig. 1(e)-(f).  We do not consider zippering because it is a slower, less likely process, which presumably depends on the dynamics of the cross-linkers. We also do not consider depolymerization as it occurs on a much slower time scale than polymerization.   

Both tip-tip bundling and capping occurs at bundle/filament tips. We also assume that branching occurs at the growing bundle/filament end as well.  The assumption of branching at the tip underestimates the effects of branching since side branching has been rather well-established experimentally~\cite{volkmann}. 
However, there is also a known aging effect where Arp2/3 is found to bind 
preferentially to ATP-actin in filaments~\cite{vavylonis}. In other words, branching is biased towards 
the growing end, which has also been observed experimentally~\cite{amann,ichetovkin,fujiwara}.  These assumptions allow us to introduce the variable, $\rho_i({\bf x},t)$, the density of growing tips in a bundle of size $i$ at position ${\bf x}$ and time $t$. Tip-side bundling can be incorporated into the model by introducing an average growing length of bundles.  We will address this extension in the next section.

How does one describe the dynamics of the growing tips? Since actin is a semiflexible polymer with a persistence lengthscale of about 17 microns, and the experiments indicate a star size on the order of 10 microns, the tip itself does not diffuse, but rather the angle of the tip with respect to some fixed axis does. In other words, over the lengthscale of the experiment, the motion of the filament tips is essentially ballistic. 

Putting all of the above ingredients together yields an equation of motion for $\rho_1$,
\begin{equation}
\partial_t\rho_1({\bf x},t)=-v_p{\bf \nabla} \rho_1({\bf x},t)+{\mathcal L}({\bf x})\rho_1({\bf x},t)+b\rho_m^2(t)\sigma({\bf x},t)\sum_{i=1}g_i\rho_i({\bf x},t) -c\rho_1({\bf x},t)+2c\rho_2({\bf x},t)- 2\lambda_{1,1}({\bf x},t) \rho_1^2({\bf x},t),
\end{equation}
where ${\mathcal L}({\bf x})$ accounts for the angular diffusion of the tip and $v_p$ is the polymerization velocity. We have allowed for spatial variations in the density of Arp2/3 denoted by $\sigma({\bf x},t)$. The density of G-actin is denoted by $\rho_m({\bf x},t)$,   It is known that the nucleation of a branch at rate $b$ requires a dimer of G-actin monomers hence the quadratic dependence on $\rho_m({\bf x},t)$. The $g_i$ factor is to account for any geometric factors related to the dimensionality of the system and the size of the bundle. Note that the branching depends on the total concentration (or number of) bundles as opposed to the total concentration of filament tips. Finally, capping 
occurs at some uniform rate $c$ and the bundling of two single filaments occurs at some rate $\lambda_{1,1}({\bf x},t)$.  

Since we have introduced the concentrations of Arp2/3 and G-actin, we now construct their respective equations of motion. Arp2/3 diffuses and get taken up into branches.  Therefore, the density of Arp2/3 is given by 
\begin{equation}
\partial_t\sigma({\bf x},t)=D_{Arp2/3}\nabla^2\sigma({\bf x},t)-b\rho_m^2(t)\sigma({\bf x},t)\sum_{i=1}g_i\rho_i({\bf x},t),
\end{equation}
where $D_{Arp2/3}$ is the diffusion coefficient for Arp2/3. The concentration of G-actin is governed by
\begin{equation}
\partial_t \rho_m({\bf x},t)=D_m\nabla^2\rho_m({\bf x},t)-k_p\sum_{i=1}i\rho_i({\bf x},t)\rho_m({\bf x},t)-b\rho_m^2({\bf x},t)\sigma({\bf x},t)\sum_{i=1}g_i\rho_i({\bf x},t),
\end{equation}
where $k_p$ is the polymerization rate and $D_m$ is the diffusion constant for G-actin. Note that the polymerization depends on the total filament density (or number) since each filament within the bundle polymerizes to extend the entire bundle. 

To complete the description of the system, the equation for $\rho_i({\bf x},t)$ with $i>1$ is 
\begin{equation}
\partial_t\rho_i({\bf x},t)= -v_p{\bf \nabla} \rho_i({\bf x},t)+{\mathcal L}({\bf x})\rho_i({\bf x},t)+\sum_{k=1}^{i-1}\lambda_{k,i-k}\rho_k({\bf x},t)\rho_{i-k}({\bf x},t)-2\sum_{k=1}\lambda_{i,k}\rho_i({\bf x},t)\rho_k({\bf x},t)+(i+1)c\rho_{i+1}({\bf x},t)-ic\rho_{i}({\bf x},t).
\end{equation}
As above, the first term on the right accounts for the ballistic motion, the second term accounts for the angular diffusion of the tip. The third term describes those bundles of size and 
$k$ and $i-k$ coming together to become a bundle of size $i$, while the fourth term describes those bundles that were of size $i$ at the previous 
time step now becoming larger via bundling. We will compute $\lambda_{i,k}$ in the next section within a mean field approach. 
     
The above equations constitute a general scenario. For the remainder of this work, we will focus on a quasi-two-dimensional system---quasi-two-dimensional in that the filaments can overlap one another and so we do not take into account excluded volume. In the initial experiments, stars were detected in regions of the solution with more spread~\cite{vignjevic}, i.e. thinner regions.  Hence our reason for studying a quasi-two-dimensional geometry. In this particular geometry, we assume that the Arp2/3 initiates a branch in the plane via binding to either outer edge of a bundle tip, hence $g_i=2$. We note, however, that the more recent experiments were conducted in three-dimensional solutions.  We will address the application of this model to three-dimensions in the discussion section. 

\section{Mean field theory}
Before presenting the results of our kinetic simulations, we introduce a set of simplifications of the two-dimensional model to determine the scaling form of the bundle size distribution, ultimately.  This result can be tested in the kinetic simulations. The first simplification allows us to compute the bundling rates explicitly. A second simplification allows for comparison with previous work. A final simplification reduces the model to a single set of coupled differential equations.
 
The first simplification involves taking the spatially-averaged limit of Eqns. (1)-(4).  In mean field, Eqns. (1)-(4) become  
\begin{equation}
\frac{d\rho_1(t)}{dt}= 2b\rho_m^2(t)\sigma(x,t)\sum_{i=1}\rho_i(t) -c\rho_1(t)+2c\rho_2(t)- 2\lambda_{1,1} \rho_1^2(t),
\end{equation}
\begin{equation}
\frac{d\sigma(t)}{dt}=-2b\rho_m^2(t)\sigma(t)\sum_{i=1}\rho_i(t),
\end{equation}
\begin{equation}
\frac{d \rho_m(t)}{dt}=-k_p\sum_{i=1}i\rho_i(t)\rho_m(t)-2b\rho_m^2(t)\sigma(t)\sum_{i=1}\rho_i(t),
\end{equation}
\begin{equation}
\frac{d\rho_i(t)}{dt}= \sum_{k=1}^{i-1}\lambda_{k,i-k}\rho_k(t)\rho_{i-k}(t)-2\sum_{k=1}\lambda_{i,k}\rho_i(t)\rho_k(t)+(i+1)c\rho_{i+1}(t)-ic\rho_{i}(t)\,\,\,\,\,\,\,{\rm for}\,\,\,\,\,\,i\ge 2.
\end{equation}
  
\subsection{Tip-tip versus tip-side bundling}

We can now compute $\lambda_{i,k}$, which is the tip-tip collision rate for two individual ballistically moving filament tips in two dimensions. It is given by $\lambda_{1,1}^{tt}(t)=(\theta_c/180^{\circ})(2d_G+d_X)v_{rel}(t)$, where $d_G$ is a G-actin diameter and $d_X$ is a crosslinker diameter, $v_{rel}(t)$ is the relative velocity between two growing filaments, and $(\theta_c/180^{\circ})$ accounts for the small angle intersection required for bundling.  The relative velocity is $\frac{4}{\pi}\frac{\rho_m(t)}{\rho_m(0)}v_p$, where $v_p$ is the polymerization velocity. Now that we have calculated $\lambda_{1,1}^{tt}$, the three-dimensional cross-section for a bundle of size $i$ projected into the plane scales as $(d_G+d_X)\sqrt{i}$. Therefore, the tip-tip collision/bundling rate for two bundles of size $i$ and size $k$ coalescing is $\lambda_{i,k}^{tt}=\frac{4}{\pi}\lambda_0(t)(\sqrt{i}+\sqrt{k})$ with $\lambda_0(t)$ approximated as $(\theta_c/180^{\circ})(d_G+d_X)\frac{\rho_m(t)}{\rho_m(0)}v_p$.  Note that we have neglected the cross-linker diameter joining the two bundles, and we have assumed that the cross-linkers can reorient arbitrarily large bundles.  As for experimentally relevant numbers for $\lambda_{i,j}^{tt}$, we use $d_G=0.003\,\mu m$, $d_X=0.007 \,\mu m$, $v_p=0.3\,\mu m/sec$, $\theta_c=20$ degrees~\cite{yang,yu} and $\rho_m(0)=10\,\mu M$. The $d_X$ we have chosen is roughly the size of fascin.  The diameter of alpha-actinin is approximately 30 nm.   Both fascin and alpha-actinin were used in the initial experiments.

As for our second simplification, we set the concentration of Arp2/3 to be constant. In spatially homogeneous system, we cannot study the spatial localization of Arp2/3 presumably required for star formation. Since branches also require G-actin to extend, we keep track of the time-dependence of the G-actin so as to observe an upper limit on the bundle size distribution, i.e. that the bundle size distribution eventually stops evolving, as is indicated in the experiments. We now have  
\begin{equation}
\frac{d\rho_i(t)}{dt}= \sum_{k=1}^{i-1}\lambda_{k,i-k}^{tt}(t)\rho_k(t)\rho_{i-k}(t)-2\sum_{k=1}\lambda_{i,k}^{tt}(t)\rho_i(t)\rho_k(t)+\delta_{1,k}\bar{b}\rho_m^2(t)\sum_{i=1}\rho_i(t)+(i+1)c\rho_{i+1}(t)-ic\rho_{i}(t)
\end{equation}
\begin{equation}
\frac{d\rho_m(t)}{dt}=-k_p\sum_{i=1}i\rho_i(t)\rho_m(t)-\bar{b}\rho_m^2(t)\sum_{i=1}\rho_i(x,t),
\end{equation}
where $\bar{b}=2b\sigma(0)=2b\sigma$. Equations 9 and 10 describe an aggregating system with a time-dependent aggregation rate, a time- and concentration-dependent source term, and a size-dependent death rate via capping. As G-actin polymerizes into bundles, bundle extension slows.  Once the G-actin is used up, the bundles cannot extend any further to collide with other growing bundles and the system stops evolving.  
 
We study Eqns. (9) and (10) numerically.  To do this, we first discuss units.  We normalize any density by $\rho_1(0)=0.002\,\mu M$, the initial density of filaments. This initial density is obtained using a G-actin concentration of $10 \,\mu M$ and a dimerization rate of $3 \times 10^{-6}\mu M^{-1} s^{-1}$ allowing the solution to dimerize over the course of about 100 seconds.  What number density does $0.002\,\mu M$ correspond to? Since we are assuming a two-dimensional geometry for the collision process and one $\mu M$ is roughly equivalent to 600 particles per $\mu\,m^3$, the effective two-dimensional concentration is $0.002\,\mu M \,(600 \,tips/\mu m^3) (0.1 \,\mu m)=0.12 \,tips/\mu m^2$, where $0.1 \,\mu m$ is the estimated thickness of the star.  We use this number density when normalizing the two-dimensional collision rate. Now we normalize the time units.  Since the polymerization of the filaments is the fastest time scale, using $k_p=10 \,\mu M^{-1} s^{-1}$ and $\rho_1(0)=0.002 \,\mu M$, we arrive at a time scale of $50\,s$, which is in rough agreement with the experiments~\cite{haviv}.

Using the experimentally relevant parameters listed above, as well as $b=0.1 \,\mu M^{-3} s^{-1}$~\cite{carlsson}, we numerically integrate the mean field system of bundling equations via the Euler method. We impose an upper cutoff, $I$, on the number of bundles and check that the step size of 0.001 is small enough such that $\rho_i(t)>0$ as was done in Ref.~\cite{kang}.  We have checked that our results are essentially independent of the step size and $I$. With this method, we find that for the relevant concentrations of Arp2/3, the tip-tip bundling rate is not large enough to accommodate the excess of single filaments, at least for the initial condition of $\rho_i(0)=\delta_{i,1}$ in normalized units, the monodisperse initial condition. More precisely, the average bundle size decreases from 1.05 to 1.01 as the concentration of Arp2/3 is increased from 0.005 $\mu M$ to $0.05\,\mu M$. Thus, the tip-tip bundling rate is not large enough as compared to the source rate to induce substantive bundling within the fixed timescale of the experiment.  

To arrive at a finite typical bundle size within the time scale of the experiment for the monodisperse initial condition, we include another bundling process---tip-side bundling.  Such a process occurs and, presumably, it occurs at a higher rate then the tip-tip bundling process due to the added length dependence. The tip-side bundling rate for a bundle of size $i$ coalescing with another bundle of size $k$ is given by $\lambda_{i,k}^{ts}=\lambda_0(t)(\sqrt{i}+\sqrt{k})L(t)$, where $L(t)$ is average length of the ``side'' bundle. In units of the monomer size, the equation for $L(t)$ is 
\begin{equation}
\frac{dL}{dt}=k_p\rho_m(t).
\end{equation}
As for $L(0)$, the average length of the initial seeds is not necessarily well-characterized. The dimerization rate used above assumes that $L(0)=2$, therefore, that is what we implement. 

For tip-tip bundling, the average bundle size decreases with increasing $\sigma(0)$ due to the uptake of G-actin into branching. However, for tip-side bundling there is an added effect.  Namely, since the length of filaments is dependent on the concentration of G-actin, the more Arp2/3 there is, the smaller the average length of filaments, the smaller the tip-side bundling rate resulting in a smaller bundle size typically.  This dependence is in accordance with a molecular dynamics simulation based on the later star experiments~\cite{ideses}.  Certainly, this trend is observed for large Arp2/3 concentration where stars no longer form since the filaments remain short and only a dendritic cloud forms.  However, in the experiments, it has been observed that a critical Arp2/3 concentration is needed to initiate star formation.  This is presumably because there has to be enough filaments around to bundle.  In our mean field model, this latter dependence can be accounted for by decreasing the initial filament concentration to such a level that the typical time scale becomes slower and slower such that depolymerization has to be taken into account.  Since we do not take depolymerization into effect, we presumably we will not be able to account for a threshold concentration of Arp2/3 in bundle formation. 

The length dependence of the tip-side bundling rate is not as straight forward as above. For small angle collisions to occur, the entire length of a filament/bundle is not necessarily available for bundling.  To estimate the effective length for bundling requires knowledge of the spatial distribution of the initial individual filaments and of the Arp2/3. In the absence of Arp2/3, for a distribution of initial single filaments whose radial position is much smaller than the final typical filament length, tip-tip bundling is more likely to occur than tip-side bundling.  However, for a more broadened distribution of initial seeds, the ratio of tip-side to tip-tip collisions will increase. The spatial distribution of the Arp2/3 must also be taken into account since the more broad the distribution of Arp2/3, the ratio of tip-side to tip-tip collisions may potentially increase.   Kraikivski and collaborators~\cite{kraikivski} incorporate this effect into the tip-side bundling rate by decreasing the critical bundling angle.  Since spatial information is crucial for estimating the available length for bundling, we merely study $\rho_i$ as a function of $\theta_c$. 


Now that we have fully defined the various parameters in mean field, we integrate Eqs. 9 and 10 with tip-side bundlingand neglect tip-tip with $c=0$ for the remainder of the section. When we take into account tip-side bundling, we obtain average bundle sizes larger than unity within the time scale of the system.  

In Figure~\ref{fig:mf.bundlesizedist.arp.loglin}, we plot the final bundle size distribution for different concentrations of Arp2/3. To plot the bundle size distribution, or the probability of having a bundle of a particular size, we plot $\frac{i\rho_i}{\sum_{i=1} i\rho_i}$ as a function of $i$. The denominator is needed for normalization since the mass density is not conserved.  Plotting the final bundle size distribution on a log-linear plot, we see that eventually the distribution falls off exponentially with bundle size with the decay depending on the concentration of Arp2/3 (see Fig.~\ref{fig:mf.bundlesizedist.arp.loglin}).  It is this decay that sets the average length.  
 
We also plot the average bundle size as a function of Arp2/3 concentration (Fig.~\ref{fig:mf.averagebundlesize.vs.arp-a}). As expected in mean field with no depolymerization, an increase in Arp2/3 concentration decreases the average bundle size due to the faster decay of G-actin. In addition, the faster G-actin gets soaked up into branches, the smaller the average length of filaments/bundles, thereby, again, decreasing the bundling rate and, hence, the average bundle size. We also note that bundles are able to form in the absence of Arp2/3 with this initial $\rho_1(0)$.

In Figure~\ref{fig:mf.averagebundlesize.vs.thetac-b}, we plot the average bundle size as a function of $\theta_c$.  This parameter is an attribute of the cross-linker and so can be studied by modifying/changing the cross-linker. The bundling rate is proportional to $\theta_c$, so an increase in $\theta_c$ increases the average bundle size. However, the dependence is complicated by the fact that the source term depends on the total number of bundles such that the more bundling, the less branching, the longer the filaments thereby increasing the average bundle size even more so. 

\underline{\it Comparison with previous work}: To make a more direct comparison with the work of Kraikivski and collaborators~\cite{kraikivski}, we simplify the branching/source term to $\hat{b}\rho_m(t)\sum_i\rho_i(0)=\hat{b}\rho_m(t)$, where $\hat{b}=2b\sigma\rho_m(0)$. With this simplification, we obtain exact expressions for $\rho_m(t), L(t)$, and $M(t)$, the total mass density.  In particular, we have
\begin{equation}
\rho_m(t)=\frac{\tilde{k}_p D_1}{\hat{b}}({\rm sech}(\frac{1}{2}\tilde{k}_p \sqrt{2 D_1}(t+D_2)))^{-2}
\ee
\begin{equation}
M(t)=\sqrt{2D_1}\tanh(\frac{1}{2}\tilde{k}_p \sqrt{2 D_1}(t+D_2))-\frac{\hat{b}}{\tilde{k}_p},
\end{equation}
with $D_1$ and $D_2$ determined by initial conditions and $\tilde{k_p}=k_p\rho_1(0)$. Once we know $M(t)$, we also know $L(t)$. Replacing $L(t)$ with its long-time value, we analyze 
\begin{equation}
\frac{d\rho_i(t)}{dt}= \lambda_0(t)L(\infty)\sum_{k=1}^{i-1}(\sqrt{k} + \sqrt{i-k})\rho_k(t)\rho_{i-k}(t)-2\lambda_0(t)L(\infty)\sum_{k=1}(\sqrt{k}+\sqrt{i})\rho_i(t)\rho_k(t)+\delta_{i,1}\hat{b}\rho_m(t).
\end{equation}
Reparameterizing time with 
\begin{equation}  
\frac{d\tau}{dt}=\lambda_0(t),
\end{equation}
we arrive at
\begin{equation}
\frac{d\rho_i(\tau)}{d\tau}= L(\infty)(\sum_{k=1}^{i-1}(\sqrt{k} + \sqrt{i-k})\rho_k(
\tau)\rho_{i-k}(\tau)-2\sum_{k=1}(\sqrt{k}+\sqrt{i})\rho_i(\tau)\rho_k(\tau))+\delta_{i,1}b'
\end{equation}
with $b'\propto \hat{b}$. 

Numerical simulations of this further simplified model show that tip-tip bundling alone is not enough to produce substantive bundle sizes and so we do not expect the tip-tip bundling to be the dominant mechanism of bundling.  Kraikivski and collaborators~\cite{kraikivski} reach a different conclusion.  We suspect this difference is due to the fact that our time scale involves the polymerization rate, the tip-branching rate, and the initial concentration of individual filaments.  The time scale computed by Kraikivski and collaborators~\cite{kraikivski} does not appear to involve the initial concentration of individual filaments. However, since Arp2/3 must branch on pre-existing filaments, we expect the time scale to depend on this parameter.   To arrive at a time scale comparable to the experiments, the branching rate for Kraikivski and collaborators is several orders of magnitude smaller than ours such that their branching/source rate is more comparable to their tip-tip rate so larger bundles are able to form in their model.

\subsection{Scaling form for bundle size distribution}

The numerical results of the previous subsection suggest an exponential form for the bundle size distribution, at least for higher concentrations of Arp2/3. Interestingly, an exponential tail has been measured for an actin-fascin system~\cite{haviv2}. While a finite of amount of G-actin accounts provides for an upper bound on the time evolution of the system and, therefore, determines the radial extent of the system, we expect the bundle size distribution to stop evolving before all of the G-actin polymerize (in the absence of zippering).  Therefore, we approximate the G-actin concentration to be constant. Such a system can be constructed by supplying the homogeneous system with G-actin at a rate that equals the loss of G-actin to polymerization and branching.

If we assume a constant G-actin concentration, then $L(t)$ increases linearly in time. We are free to reparameterize time using
\begin{equation}
\frac{d\tau}{dt}=\lambda_0(t)L(t),
\end{equation}
with $\lambda_0(t)$ normalized appropriately such that $\tau\sim t^2$.  Then equation for the density of bundles sizes becomes
\be
\frac{d\rho_i(\tau)}{d\tau}= \sum_{k=1}^{i-1}(\sqrt{k} + \sqrt{i-k})\rho_k(\tau)\rho_{i-k}(\tau)-2\sum_{k=1}(\sqrt{k}+\sqrt{i})\rho_i(\tau)\rho_k(\tau)+\delta_{k,1}g(\tau),
\ee
with $g(\tau)=\frac{\bar{b}\rho_m^2(0)}{\sqrt{2\lambda_0(0)\tau}}N(\tau)$, where $N(\tau)=\sum_i\rho_i(\tau)$.  A simpler form of this equation was studied by Davies and collaborator~\cite{davies} where the time-dependent source term did not depend on the density. We extend those results here.

Before doing so, in the absence of the source term ($ns$), there is no exact solution to date. However, since the bundle size dependence of the collision rate is homogeneous of degree $p$, i.e. $\sqrt{i}+\sqrt{k}=k^p(1+(k/i)^{1/2})$ with $p=1/2$, there has been much study of the scaling form for $\rho_i(\tau)$~\cite{leyvraz,vandongen}. The dynamic scaling hypothesis~\cite{leyvraz,vandongen} predicts that 
\begin{equation}
\rho_i^{ns}(\tau)\sim \frac{1}{s(\tau)^2}\Psi(\frac{i}{s(\tau)}), 
\end{equation}
at large times with $s(\tau)\sim \tau^{2}$. The $s^{-2}$ prefactor enforces mass conservation and the scaling of $s$ with $\tau$ can be derived by using Eq. 19 in the continuum limit of the kinetic aggregation equation given by
\be
\partial_{\tau}\rho(y,\tau)=\int_0^y dz (z^{1/2}+(y-z)^{1/2})\rho(z,\tau)\rho(y-z,\tau)-2\int_0^\infty dz (z^{1/2}+y^{1/2})\rho(y,\tau)\rho(z,\tau).
\ee
Moreover, $\Psi(x)$ is expected to fall off at least exponentially at large $x$, and for small $x$, $\Psi(x)\sim x^{-\gamma}$ with $1<\gamma<3/2$~\cite{cueille}. The lower bound enforces the normalizability of the distribution.  The upper bound comes from asymptotic analysis of the kinetic aggregation equation using the dynamic scaling hypothesis in the limit of small $x$. Tighter bounds can be computed~\cite{cueille}.  We refer the reader to the excellent review by Leyvraz~\cite{leyvraz} where this analysis is presented in detail. 

And now for the analysis in the presence of branching. For small bundle sizes and long enough times, one would expect the aggregation terms and source term to balance one another.  In other words, one would expect a pseudo steady state ($pss$) to emerge with 
\be
\rho_i^{pss}\sim \sqrt{g(\tau)}i^{-\Delta},
\ee
with $\Delta=7/4$, which is obtained from the constant source case\cite{krapivsky}. For intermediate bundle sizes and long enough times, one expects a self-similar solution of the form 
\be
\rho_i(\tau)\sim \tau^{-7/4}\bar{\Psi}(i/\tau^{1/2}), 
\ee
where we used (1) the conservation of mass in the presence of a time-dependent source, or $\sum_{i=1}  i\rho_i(\tau) \sim \int d\tau N(\tau)/\tau^{1/2}$, and (2) the scaling form for $\rho_i(\tau)$ consistent with the source-free aggregation equation with this particular form of the collision rate, or $\rho_i(\tau)\sim \tau^{-(1+\frac{3}{2}q)}\bar{\Psi}(i/\tau^{q})$.  For this intermediate range of $i$, a scaling collapse is demonstrated in Figure~\ref{fig:mf.scalingcollapse.constantrhom-a}.  With this scaling for $\rho_i(\tau)$ in the intermediate region, the integrated newly introduced mass eventually decays with time and the system approaches the constant mass (no source) case with $\rho_i(\tau)$ scaling as $\rho_i^{ns}(\tau)$.  See Figure~\ref{fig:mf.scalingcollapse.constantrhom-b} for a scaling collapse for this large bundle size region solution.  

While we may not observe all of these regimes in finite-dimensional simulations of such a system, since presumably there is an upper bound on the bundle size, based on the above analysis, it is reasonable to expect that  
\be
\rho_i \sim i^{-\Delta} \exp(-i/i_{ch}),
\ee
where $i_{ch}$ is some characteristic bundle size that depends on the amount of Arp2/3, for example, and $\Delta\approx 7/4$.
  
\section{Kinetic Simulations} 

Now that we have a prediction for the scaling form for the bundle size distribution in the spatially homogeneous system, let us test what aspects of that prediction will persist in the kinetic simulations with spatial degrees of freedom. We will also measure the local orientation of the filaments to determine if indeed a star has formed as opposed to a network of bundles, which are not differeniated in the analysis above. To accomplish these tasks, we perform kinetic aggregation simulations in quasi-two-dimensions. More specifically, we allow for a third dimension only in that we allow growing filaments/bundles to cross each other.  Furthermore, we not only keep track of small angle cross-links that leading to bundling but also large angle cross-links as well. These large angle cross-linkers do not reorient the filaments, but account for connected clusters of filaments. These connected clusters will become important when including the finite-size of a bead embedded in the system to closely model the initial experiments. We do not incorporate dynamics of the linkers.  Once two filaments are bound, they remain bound for the time scale of the experiment. 

\subsection{Algorithm}
To simulate the actin-Arp2/3-cross-linker system we keep explicitly keep track of the filaments and the Arp2/3, but not the G-actin and the cross-linkers. In other words, the concentration of G-actin and cross-linkers are parameters, while the positions of the actin in filaments are updated and recorded as are the positions of the Arp2/3. Explicit tracking of the Arp2/3 allows us to test the assumption that localization of the Arp2/3 is required for star formation since Arp2/3 promotes dendritic growth as opposed to oriented, bundled growth.  As for the constant concentration of G-actin, in a spatially homogeneous system, the filaments bundle indefinitely. However, in a system with spatial degrees of freedom, the bundling rate acquires spatial dependence such that as bundles orient as they extend outward from the center of the system, the bundling decreases towards the perphiery of the system in even the presence of G-actin. We expect this spatial dependence, and not the G-actin concentration, to limit the bundle size in the absence of zippering.  Of course, a finite amount of G-actin is required to limit the radial extent of the star. Since previous numerical studies have focused on this aspect, we will not focus on it here~\cite{ideses} and assume that filaments grow at the same rate near the center of the system.  As for the constant concentration of cross-linkers, the initial experiments found cross-linkers both the dendritic and bundled regions thereby justifying our simplification. \\
{~}\\
We perform the following steps:\\
(1) Initialize $N_F$ single filaments at some random position with the plus end pointing at some random angle uniformly distributed between $[0,2\pi]$. The random positions are chosen uniformly  within an annulus whose inner radius is $R_b$, the radius of the bead, and whose outer radius is $R_b+R_i$.\\
{~}\\
(2) At each simulation time step, a filament grows by 100 nanometers, or approximately 30 monomers.  This compression of data storage will allow us to keep track of percolation (connected cluster) properties of the cross-linked network over length scales of the experiment. In addition, at each simulation time step, the angle of the segment is varied by a Gaussian random angle with a variance of 3 degrees with respect to the angle at the previous simulation time step to approximate a persistence length of tens of microns.  When $R_b>0$, a filament cannot grow in the bead region unless the entire filament and the filaments connected to it via small-angle and large-angle cross-links, the entire cluster, can be pushed backward/outward away from the bead.  In other words, the entire connected cluster ratchets off the bead. To keep track of the clusters of cross-linked filaments, small angle and large angle, we implement the Hoshen-Kopelman algorithm~\cite{hoshen}. \\
{~}\\
(3) After updating the growth of each filament, a check is done to determine if any two new segments of two growing filaments intersect at an angle less than $\theta_c$ or are within a distance $d_X+d_G$ of another filament and are about to intersect at an angle of less than $\theta_c$. If so, then the filament in the smaller bundle is reoriented towards the filament in the largest bundle.  A check is also made if the new segment intersects with any older segment of another filament.  If so, the new segment realigns with the older segment (provided the intersection angle is less than $\theta_c$) and a bundle is formed. For both checks, new with new and new with old, if the intersection angle is greater than the two filaments become ``permanently'' connected via a cross-linker, but no realignment is made.\\   
{~}\\
(4) As for the Arp2/3, the position of $N_A$ Arp2/3 is initially randomized in the same way that the initial filament seeds are. The Arp2/3 subsequently diffuses with a diffusion constant corresponding to approximately 1 $\mu m^2/s$. After updating the positions of the Arp2/3 in parallel, a check is made to determine whether an Arp2/3 is within a distance $d_A$ of a filament.  If so, a new filament is nucleated 70 degrees from the growing end of the filament and the position of the Arp2/3 is recorded.\\ 
{~}\\
(5) Steps (2) through (4) are repeated until all the Arp2/3 binds.  Then, the positions of the growing ends of the filaments are recorded. Also, the local orientation of the filaments as a function of distance from the center is measured via the following method:\\
Two segments are chosen at random, if the two segments are within a distance $d_f$ of each other, then the cosine of the angle between the two segments is calculated and binned according to the distance from the center. \\   
{~}\\
(6) Since we are interested in how bundles at the extremities of the system emerge from the center of the system, those filaments whose growing end point outwards and exceed a radius $R_{cut}$ and are tagged and continue to grow via (1)-(4) until the bundle size distribution, which is computed at each time step, stops evolving. The local orientation of the filaments is measured again further out.  A final bundle size distribution is also outputted.  The final bundle size distribution will depend on $R_{cut}$.  We will use this dependence to say something about the nucleation of bundles from the center of the system.\\
{~}\\
(7) (1)-(6) is repeated 1000 times for sample averaging.\\
{~}\\
Two snapshots of the simulations, one without a bead and one with a bead, are depicted in Figs.~\ref{fig:snapshot-a} and ~\ref{fig:snapshot-b}.

We also note for computational simplicity we do not explicitly increase the bundle size when filaments bundle.  The reorientation of the individual filaments for bundle lengths smaller than the persistence length is sufficient to keep the two filaments aligned.  In other words, we did not capture the $\sqrt{i}$ bundling rate dependence explicitly, however, the bigger the bundle, the more likely it is to collide with another bundle due to the increase in local filament density. For (6), we also decrease the amount of angular diffusion of the tips by $1/i$, where $i$ is the bundle size.  This is to capture the appropriate persistence length dependence for bundles of filaments. 
 
\begin{table}[h]
\caption{Table of Parameters}
\centering
\begin{tabular}{ l  c  r }
\hline\hline 
Parameters &  Variables & Values\\ \hline
Polymerization rate & $k_p$ & $10$ $\mu M^{-1} s^{-1}$\\
Branching rate & $b$ & $0.1\,\,\mu M^{-3}s^{-1}$ \\
Capping rate & $c$ & $0-1\,s^{-1}$\\
G-actin concentration (when fixed) & $\rho_m$ & $\sim 10\,\,\mu M$\\
Arp2/3 concentration (when fixed) & $\sigma$ & $0-10\,\, nM$\\
Diffusion constant for Arp2/3 & $D_{Arp2/3}$ & $1\,\,\mu m^{2} s^{-1}$\\
Critical bundling angle & $\theta_c$ & $0-30^{\circ}$\\
Diameter of G-actin & $d_G$ & $3\,nm$\\
Diameter of cross-linker & $d_X$ & $7-17\, nm$\\
Arp2/3-filament binding distance & $d_{A}$ & $0.01\,\mu m$\\
Local orientation distance & $d_f$ & $0.01-0.03\,\mu m$\\
Radius of the bead & $R_b$ & $0.5\,\mu m$\\
Max. rad. of initial filament seeds from center - $R_b$& $R_i$ & $0.4-1.5\,\mu m$\\
Radius for bundle size determination & $R_{cut}$ & $0.75-0.95\,\mu m$\\
Number of initial filament seeds & $N_F$ & $100-500$\\
Number of Arp2/3 & $N_A$ & $250-650$\\
\end{tabular}
\end{table}    

\subsection{Results}

\subsubsection{Absence of bead: $R_b=0$}

\underline{\it Bundle size distribution}: 

In Fig.~\ref{fig:2d.star.nobead.bundlesizedist.dynamic} we plot the logarithm of the bundle size distribution for $\theta_c=30^{\circ}$.  As for the other parameters, $R_i=0.6 \,\mu m$, $N_F=500$ and $N_{A}=250$.  And, unless otherwise specified, $d_X=0.017\, \mu m$, $d_G=0.003\, \mu m$, $d_{A}=0.01\, \mu m$, $d_f= 0.03 \,\mu m$, and $R_{cut}=R_b+0.75\, \mu m$.  Note that $d_X$ is slightly larger than in mean field, at least for the bulk of the simulations. Also note that the concentration of initial seeds is several orders of magnitude higher than in mean field depending on the thickness of the star. 

From Fig.~\ref{fig:2d.star.nobead.bundlesizedist.dynamic} find that the bundle size distribution approaches a limit as time proceeds. In other words, the collision rate decays substantially with time even in the presence of G-actin over the lengthscale of approximately 5 microns. This is because once the bundles form near the center of the system, collect together and polymerize outward in an oriented fashion, the spacing between bundles increases, therefore, the collision rate decreases. Bundles are less likely to merge at the system's edge. 

We investigate the effect of $\theta_c$ on bundle formation.  If all other parameters are known, measurement of the bundle size distribution allows one to determine the amount of reorientation by the cross-linker---a microscopic detail. From the mean field kinetic aggregation calculations in Section IIIB, the bundle size distribution is exponential with an algebraic correction.  See Eq. 23. In quasi-two dimensions, at least for $\theta_c>10^{\circ}$, the data suggests a similar form, but the algebraic correction is a power law with exponent approximately $\Delta-1\approx 0.6$ as opposed to $\Delta-1=0.75$ computed in Sec. IIIB. (The bundle size distribution contains an extra factor of the bundle size $i$ in comparison to $\rho_i$).  To see this algebraic correction more clearly, we also plot the log-log of Fig.~\ref{fig:2d.star.nobead.bundle1-a} in Fig.~\ref{fig:2d.star.nobead.bundle1-b}. The exponent obtained in two-dimensions may be in slightly better agreement with the one found for a constant aggregation kernel with a constant source term~\cite{leyvraz}, which is $\Delta-1=0.5$. This is reasonable given that we have not explicitly included the size of the bundle in the simulation.  Work by Connaughton and collaborators~\cite{connaughton} have investigated the role of spatial fluctuations for a system of diffusing, coalescing particles in the presence of a constant source term and found that the exponent depends on the spatial dimension for $d<=2$.  It would be interesting to extend this work to the semiflexible case to more accurately determine analytically the algebraic correction in finite-dimensions. Furthermore, it seems that the spatial distribution of the Arp2/3, which was obviously not accounted for in the mean field theory, does not alter the form of the bundle size distribution from mean field.

We have also measured the bundle size distribution as a function of concentration of Arp2/3 by changing $N_A$ but keeping $N_F+N_A$ fixed. Svitkina and collaborators~\cite{svitkina2} argue that Arp2/3 provides enough filaments for a broad, dendritic base from which filaments can emerge and bundle to form larger bundles than in the absence of Arp2/3.  This concept of a larger base leading to larger bundles is a natural one and is supported by the fact that while it is less likely to have smaller bundles the larger the Arp2/3 concentration, the more likely it is to have larger bundles thus increasing the average bundle size overall. See Fig.~\ref{fig:2d.star.nobead.bundle.arp}. The average bundle size is $15.50\pm0.32$ for $N_A=450$ and $18.46\pm0.40$ for $N_A=650$.  In contrast with mean field, here, an increase in the concentration of Arp2/3 increases the average bundle size. 

We observe a larger base leading to larger bundles concept when we study the bundle size distribution as a function $R_{cut}$.  The larger $R_{cut}$, the smaller the base from which filaments can emerge and bundle resulting in smaller bundles.  For example, for $R_{cut}=0.85\,\mu m$ (as compared with $R_{cut}=0.75\,\mu m$), the average bundle size is $9.03\pm0.18$ and for $R_{cut}=0.95\,\mu m$, the average bundle size is $5.57\pm0.13$.  

While we find support for the convergent elongation model here, at least in the absence of depolymerization, we do not find a threshold concentration of Arp2/3 for bundling.  In addition, we find that in order to achieve bundles of reasonable size (of order 10 on average), one needs enough filaments, branched or otherwise, within a small enough region.  If we increase $R_i$ to $1.8$ microns from $0.6$ microns, then the average bundle size is $7.58\pm0.14$ as compared to $14.94\pm0.20$ (with some nontrivial density dependence in between).  Beyond $R_i=1.8$ microns, the average bundle size continues to decrease. Indeed, there is a threshold filament concentration needed for substantive bundle sizes. This threshold concentration depends on somewhat on the ratio of $N_A/N_F$ but more strongly on $R_i$.  If depolymerization is included, then we should observe a threshold concentration of Arp2/3 for substantive bundling.

{~}\\
\underline{ \it Orientation of filaments:} We also measure the local orientation of the filaments with respect to one another to determine if indeed a star has formed as opposed to an isotropic network of bundles.  First, let us address several scenarios. For an isotropic system of nonpolymerizing filaments, in the absence of bundling, the average of the cosine of the angle between nearby filaments, $\Theta$, should vanish.  For a static system of isotropically distributed bundles, for length scales on the order of the bundle cross-section, the average of the cosine of the angle between nearby filaments should be unity, but on larger length scales of order the bundle cross-section, $\Theta$ will decrease.  For a static system of aligned bundles, $\Theta$ will remain close to unity even as the length scale for nearby filaments is increased. For a static system of $70^{\circ}$ branched filaments, then $\Theta=0.3$, provided the mesh size of the system is not too small. Haviv and collaborators~\cite{haviv} observe this trend in simulations modeling their experiments.
 
What about systems with polyermization? For a system of polymerizing filaments, the filaments that grow into new regions of the system tend to be oriented in the same way, even in the absence of bundling. Otherwise, they would not be able to reach the same new regions in the first place. Therefore, $\Theta>0$ for the outer exterminities of the system irregardless of bundling. 

So, for star formation to occur, we expect $\Theta(r)$ to inrease from zero near the center of the system. Given the cylindrically symmetric initial conditions, it is more probable for filaments of opposite orientation to overlap near the center. As $r$ increases, $\Theta$ should increase to approximately $0.3$ in the dendritic region with this number being mesh size dependent. Beyond the dendritic region, towards the extermities of the system, $\Theta$ will contine to increase beyond the value of $\Theta$ for the $\theta_c=0^{\circ}$ case.  

Before presenting $\Theta$ for the same parameters as in Fig.~\ref{fig:2d.star.nobead.bundle1}, we plot $\Theta(r)$ for a system where we do not expect to observe star formation in Fig.~\ref{fig:2d.star.flux.calibrate}.   More precisely, $N_F=100$, $N_A=650$, $R_i=2\, \mu m$ and $\theta_c=0$.  We expect this system to Arp2/3 dominated such that $\Theta$ is near 0.3. The spatial distribution of the Arp2/3, denoted by $Prob_\sigma$, is also plotted. We see that the average local filament orientation modulates around 0.3. If the Arp2/3 is more localized, then we expect the average local orientation between filaments beyond the Arp2/3 region to increase beyond 0.3.  To check for this, we let $R_i=1.5 \,\mu m$ and compare with the $R_i= 2.0\,\mu m$ results.  We see that beyond 1.5 microns, the Arp2/3 decays and the average local orientation between filaments increases beyond the $R_i=2.0\,\mu m$ data.

Now we plot $\Theta(r)$ as a function of $\theta_c$ for the data depicted in Fig.~\ref{fig:2d.star.nobead.bundle1}. See Fig.~\ref{fig:2d.star.nobead.flux}. For $\theta_c=0^{\circ}$, where no bundles are formed, $\Theta(r)$ increases from zero as $r$ is increased. We do not observe a plateau at $\Theta(r)=0.3$ in the Arp2/3 region (shown later).  Even if we increase $N_A$ relative to $N_F$, due to the high concentration of filaments with $d_f=0.03\,\mu m$, we do not observe the plateau. Beyond the Arp2/3 region, $\Theta$ continues to increase, until ultimately, beyond 2 microns, $\Theta$ begins to decrease indicating that filaments begin to overlap again as they grow.   

How does $\Theta$ depend on $\theta_c>0^{\circ}$? For $d_f=0.03 \,\mu m$, it turns out that $\Theta$ is slightly smaller for $\theta_c=30^{\circ}$ in comparison to $\theta_c=0^{\circ}$ near the center of the system. This result is a bit counterintuitive, but reasonable.  The reorientation of filaments biases the system towards a network of bundles as opposed to a branch dominated system in this high density regime.  Since $d_f$ is larger than the mesh size, we are more likely to measure filaments/bundles of different orientation for $\theta_c=30^{\circ}$ than for $\theta_c=0^{\circ}$. Decreasing $d_f$ increases $\Theta$ for $\theta_c=30^{\circ}$ near the center of the system such that the $\theta_c=0^{\circ}$ and $\theta_c=30^{\circ}$ curves approach each other.  Near the periphery of the system, as $\theta_c$ increases from zero, the asymptotic value of $\Theta$ increases.  The asymptote of approximately 0.95 for $\theta_c=30^{\circ}$ certainly indicates star formation when compared to the $\theta_c=0^{\circ}$ data. \\
{~}\\
\underline{{\it Summary:}} In the absence of a bead, $\theta_c$ must be sufficiently large enough to increase the bundling rate to approach an average bundle size of approximately 10 filaments per bundle.  Furthermore, there is a threshold concentration of filaments, branched or unbranched, required for substantive bundling.  There exists a spatial contribution to the collision rate for bundling that does not exist in mean field (by construction). In addition, for star formation to occur, as opposed to an isotropic network of bundles, Arp2/3 needs to be localized with bundles/filaments polymerizing beyond the Arp2/3 region.  Only in regions where the Arp2/3 is not present can $\Theta$ can attain a value near unity. Also, reorientation is needed for $\Theta$ to approach unity.

\subsubsection{Presence of bead, $R_b=0.5\,\mu m$}

Now we investigate the effect of a finite-sized bead on the bundle size distribution and the average local filament orientation.  The bead not only acts as a steric object excluding filaments, but growing filaments can ``ratchet off'' the bead.  More specifically, a filament with its growing end toward the bead fluctuates away from the bead such that a monomer can be added. Once the monomer is added and the filament fluctuates ``back'' toward the bead, the entire filament moves backward or away from the bead.  If this filament is connected to other filaments, then those filaments move backward by the same amount as well.  So, this ratcheting effect will expand the dendritic region, at least under certain conditions. 
\vspace{0.3cm}

\underline{\it Bundle size distribution:} First, we investigate the bundle size distribution. As in the $R_b=0$ case, the bundle size distribution stops evolving substantially after the bundles have localized enough to decrease the collision rate substantially towards the extremities of the system.  See Fig.~\ref{fig:2d.star.bead.bundlesizedist.dynamic}. 

We also study the bundle size distribution as a function of $\theta_c$ in the presence of the obstacle. See Fig.~\ref{fig:2d.star.bead.bundle1}.  As in the no bead case, the larger $\theta_c$, the larger the average bundle size.  More specifically, for $\theta_c=10^{\circ}$ with $R_i=1.2\,\mu m$, the average bundle size is $1.55\pm0.01$, for $\theta_c=20^{\circ}$, it is $3.48\pm0.01$, and for $\theta_c=30^{\circ}$, we measure $7.20\pm0.08$.  If we choose $R_i=0.4\,\mu m$, then the initial densities of filament seeds are approximately the same for $R_b=0$ and $R_i=0.6\,\mu m$ case ($3.6\pi$ $\mu m^2$ versus $3.9\pi$ $\mu m^2$). Then, we have for $\theta_c=30^{\circ}$, an average bundle size of $11.23\pm0.27.$ The average bundle size is smaller when compared to the system with no bead as expected.  The obstacle of the bead makes it more difficult for filament on the other side of the system to grow and eventually meet up with growing filaments of the same orientation.

The bundle size distribution data suggests that the form of the bundle size distribution is, as in the $R_b=0$ case, a decaying exponential with a power law correction. For $\theta_c>10^{\circ}$, the exponent $\Delta-1=1.0$ in comparison to the $R_b=0$ case for same initial density of seeds. This power law correction depends on the density of initial filament seeds. By increasing $R_i$ from $0.4$ microns to $1.2$ microns we find that the exponent of the power-law correction varies from approximately $\Delta-1=0.8$ to $\Delta-1=1.4$ for $R_i=1.2\, \mu m$.  

It turns out that the average bundle size decreases with increasing Arp2/3 concentration at least for a range of $R_i$ near $1.2\,\mu m$.  This result differs from the $R_b=0$ case.  However, the observation of a threshold concentration of filaments still holds.  The bundle size distribution also depends on $R_{cut}$ as in the $R_b=0$ case.  The larger $R_{cut}$ the smaller the base of bundle and the smaller the average bundle size.  Finally, we also decrease $d_X$ to $0.007\,\mu m$ and find that the average bundle size decreases to $6.76\pm0.06$.\\

{~}\\
\underline{ \it Orientation of filaments:} Now we investigate $\Theta(r)$ in the presence of the bead.  See Fig.~\ref{fig:2d.star.bead.flux-a}. Since the filaments can no longer access the center of the system, $\Theta(r=R_b)>0$. For $r>R_b$, the ratcheting effect extends the dendritic phase of the filaments such that $\Theta(r)$ is nonmonotonic for $r<2$ microns. This nonmonotonicity is due to the ratcheting effect where branched filaments get pushed outward such that there is some ``memory'' of the randomized initial conditions further away from the bead then in the $R_b=0$ case. See Fig.~\ref{fig:2d.star.nobead.flux} for comparison.

For $\Theta(r)$ even further away from the bead, $r>2\,\mu m$, we notice a second nonmonotonicity (as compared to the $R_b=0$ case). If we plot $\Theta$ as a function of $\theta_c$, we notice that the negative slope for the edge filaments increases in magnitude with $\theta_c$. Due to the ratcheting mechanism, filaments are less ordered emerging from the isotropic region such that there is a larger spread in orientation among the bundles (and individual filaments).  This large spread in orientation from bundle to bundle eventually leads to overlap between bundles at an angle not necessarily less than $\theta_c$. Hence, one observes a decrease $\Theta$ given the overlap between different bundles.  Moreover, the obstacle of the bead prevents some growing filaments from ``meeting up with each other'' even if their seeds are close to each other and have a similar orientation.  So $\Theta(r)$ is smaller for $R_b>0$ as compared to $R_b=0$. If we increase $R_i$, we decrease the ratcheting effect and the $\Theta$ become less nonmonotonic. See Fig.~\ref{fig:2d.star.bead.flux-b}. \\
{~}\\
\underline{\it Capping:} Finally, the initial {\it in vitro} experiments investigated the effects of capping protein on the star formation.  Capping protein needed   
to be depleted by a sufficient amount in order to observe the star formation.  However, an increase in the concentration of capping protein, for small concentrations, induced further cloud growth, i.e. the radial extent of the dendritic/cloud increased~\cite{vignjevic}.  This result is counterintuitive because capping inhibits polymerization so one would think that the radial extent of the network would decrease. Eventually, 
for high enough capping protein concentration, the radial extent of the network decreases such that even the formation of 
clouds is inhibited. So, the enhancement of cloud growth is not so obvious. How 
do we explain this? 

The interplay between  the ratcheting effect of the bead and the percolation of the cross-linked filaments/bundles can explain this result.
 For small capping, filaments are longer and, therefore, have a great chance of forming large-angle cross-links with other filaments. Increasing large-angle cross-links increases the likelihood that filaments/bundles are
connected to other filaments on the opposing
side of the bead, i.e. percolation across the system is more likely. Then, shorter filaments growing
into the bead cannot ratchet backward because the
filaments on the opposing side
of the bead cannot move into the bead.  However, as capping is increased,
there is less connectivity and, hence, less chance of connectivity between two sides of the bead, thereby among allowing the ratcheting effect
to dominate and enhancing the radial extent of the network.  We observe an extreme case of this in Fig.~\ref{fig:snapshot3}.  Two clouds have been ``born'' if you will, for a capping rate of $0.6 \, s^{-1}$.  See Fig.~\ref{fig:2d.star.bead.capping} for the averaged data.  As capping is increased from zero, the position of the filament's growing tips becomes more extended. Increasing Arp2/3 will mimic this effect, provided the filament density is not too large, since filaments are shorter on average thereby decreasing the likelihood of forming large-angle cross-links. \\
{~}\\
\underline{ \it Summary:} For $R_b>0$, again, we find that the bundling rate and the filament density needs to be sufficient enough to attain average bundle sizes of order 10. The bundle size distribution has the same form qualitatively as the $R_b=0$ case.  While the bead does inhibit bundle formation to some extent, less ordered star formation still occurs with $\Theta$ smaller than the $R_b=0$ case.  The presence of the bead also leads to a nonmonotonic capping protein effect via the interplay of percolation and ratcheting.

\subsubsection{Ordered initial conditions and $R_b=N_A=0$}

Finally, we go beyond modeling the experiments and we investigate the case when bending of the actin filaments over long enough lengthscales determines the bundling rate.  This case can be studied with seeds emerging from an ordered array with the same orientation. At long enough lengthscales, the filaments will ultimately bend and collide with one another at small angles and form bundles.  See Appendix A for analysis of the bundle size distribution in this case.  The simplified analysis yields a bundle size distribution that is a nonmonotonic in the bundle size, which differs from the disordered initial condition and small length scales case. However, if we incorporate the effect of reorienting filaments via bending into the disordered initial condition model, we also find a nonmonotonic bundle size distribution with bundle size. 

We test this nonmonotonic prediction numerically for $N_F=100$.  See Fig.~\ref{fig:2d.ordered}. The seeds start $0.1$ microns apart and $d_X=0.017\,\mu m$. As $\theta_c$ increases to $30^{\circ}$, for filament lengths of $8\,mm$, the bundle size distribution starts to become nonmonotonic. In other words, the distribution begins to look qualitatively different from $N_A>0$, randomized initial conditions and shorter length scales.  We expect that such a distribution can be obtained experimentally as long as the concentration of G-actin can maintained over scales much larger than the persistence length. 

\section{Discussion}

We have extended the kinetic aggregation approach to study the morphology of an actin-bundling-protein-Arp2/3 system. Such an approach began with Smoluchowski~\cite{smoluchowski} and has been used to describe numerous systems ranging from coagulating colloids to polymers to ordinary and galactic dust to name a few~\cite{leyvraz}.  Kinetic aggregation modeling is orthogonal to the equilibrium approaches of actin bundling via electrostatics~\cite{tang}. Mechanisms for setting a typical bundle size in these scenarios range from the long-range electrostatic repulsion between like-charged filaments~\cite{ha} to the interplay between electrostatics and steric effects~\cite{henle} to a energetic penalty for chiral packing~\cite{grason}. However, these ideas are applied to solutions of fixed length actin filaments with no branching. The star system contains both polymerization and branching, suggesting a nonequilibrium approach. 

With minimal assumptions (aside from assuming the system is kinetically driven), we have computed the bundle size distribution for a system of polymerizing actin filaments that branch and bundle. 
In mean field, in contrast with Kraikivski and collaborators~\cite{kraikivski}, we find that tip-tip bundling alone is not enough to generate bundles of substantive size.  Moreover, in our quasi-two-dimensional simulations, we keep track of tip-tip (new-new) bundling versus tip-side (new-old) bundling and find for most initial conditions studied that the tip-tip bundling events are approximately four to fives lower than the tip-side events. While not as dramatic a difference as in mean field, there is indeed a difference.  

In finite-dimensions, the average bundle size is controlled by a spatial factor in the collision rate, which limits the bundle sizes in a nontrivial way as the system extends itself.  Since the limiting factor comes into play near the extermities of the system, the notion of the lambda percursor is an important one for substantive bundling to occur.  Bundles should form early on within the dendritic meshwork and extend themselves as well as collide with nearby bundles soon after emerging from the dendritic network. In other words, our results support the convergent elongation model.

The conjecture that Arp2/3 is crucial for filopodia formation is not bourne out in these results, while a threshold concentration of filaments, branched or otherwise, is, at least in the absence of depolymerization.  In recent studies of neuronal cells, it has been shown that the depletion of Arp2/3 results not only in less lamellipodia but in less filopodia as well~\cite{korobova}, again, highlighting the importance of Arp2/3.  Extending our simulation to include depolymerization as well as excluded volume interactions would be useful for further studies of the conjectured importance of Arp2/3.  The excluded volume interactions may help explain how Arp2/3 localizes, even in the absence of the bead, to form stars. One could also take a step back from the experiments and explore the role the branching angle of the Arp2/3 on filopodia formation.  Edelstein and Ermentrout~\cite{edelstein} have investigated the interplay between the branching angle and cross-linking and found that for small branching angles, an instability in the isotropic state emerges such that filaments orient along one direction and so are not able to bundle effectively at least for short filaments. One may expect the larger the branching angle, the larger the base of the lambda precursor, resulting in larger bundles. 

As for the shape of the bundle size distribution, we determine in mean field that the distribution is exponentially decaying but with a power-law correction.  The crossover between the two regimes can be tuned by varying the concentration of Arp2/3.  The quasi-two-dimensional simulations exhibit this form as well but the exponent depends on various factors such as the presence of the bead.    Moreover, in the quasi-two-dimensional simulations, increasing the concentration of Arp2/3 (keeping the G-actin concentration fixed) increases the potential base for the lambda precursors, at least in the absence of the bead, leading to larger bundle sizes on average. We begin to obtain a qualitatively different bundle size distribution for filaments emerging from an ordered array, where filaments can only collide via transverse bending.  The bundle size distribution becomes nonmonotonic in the bundle size for a large enough bundling rate.  Nonmonotinicity can also be accounted for by introducing a cutoff in the number of filaments the cross-linkers can reorient given the cost in bending energy interestingly enough. Determining whether the bundle size distribution is peaked or not should presumably be the next experimental test of the star system.  We should note bundle size distribution measurements of a network of bundles in an actin-fascin system suggest that there is a peak in the distribution of bundle sizes, though evidence for the peak is essentially one data point~\cite{haviv2}. 

Initial experiments demonstrated that star formation was more likely to occur in places where there was more spreading of the solution across the coverslip, hence the quasi-two-dimensional modeling.  However, subsequent experiments and simulations have studied star formation in three-dimensions~\cite{haviv,ideses}.  The model studied here can be extended to three-dimensions.  In the mean field theory, the parameter $g_i$ introduced in Equation 1 will now be size dependent since the Arp2/3 could branch off of the surface area of the bundle as opposed to the two outer filaments in each bundle in two-dimensions.  The larger the bundle, the more likely for Arp2/3 to branch off its surface area. This increase in the nucleation of new filaments will offset, to some extent, the decrease in the collision rates in three dimensions and presumably account for star formation in three-dimensions as is observed. In three dimensions, one expects tip-side bundling to dominate over tip-tip bundling to an even greater extent. 

While we have focused on bundle formation via kinetic aggregation in the absence of a membrane,  others have begun to study the interplay between bundles and the membrane. Atiligan and collaborators~\cite{atiligan} propose a filament-membrane interaction and show that filopodia extension rate is faster for an elastic membrane as compared to a rigid membrane.  Lan and Papoian~\cite{lan} build on the previous simulations by incorporating monomeric flow to demonstrate that rate of filopodia extension is very sensitive to the retrograde flow of monomers.  Other work has investigated the possibility of the membrane inducing bundling~\cite{fletcher}.  It would be interesting to extend our work to include a dynamic obstacle and determine its effect on bundle formation.  We should note that a more recent simulation by Zhuravlev and Papoian~\cite{zhuravlev} studies in detail the effects of the binding and unbinding of capping protein on filopodia formation and discovers large fluctuations in the filopodia lengths as compared to in the absence of capping protein.  

The interplay between ratcheting and percolation has been explored to some extent in this work.  Most percolation analysis has been reserved for ``static'' systems, where energetics dominates~\cite{carlsson1}.  For example, recent Monte Carlo simulations of a system of binary rods, one rod to model the actin filaments, the other to model the cross-linkers, have explored the percolation transition in isotropic and bundled phases~\cite{chelakkot}.  In the isotropic phase, the percolation transition resembles that of a vulcanization transition. It would be interesting to explore more thoroughly the percolation transition as a function of growing rods in the context of various phases ranging from the isotropic to dendritic to star phases as well as potentially new phases. 

Finally, star formation has recently been observed in nonequilibrium actin/cross-linker networks where polymerization is initiated~\cite{lieleg}. These stars form in the absence of Arp2/3.  We have seen this to be case in our calculations. Lieleg and collaborators~\cite{lieleg} have measured the fractal dimension of the stars to be around $1.8$.   A study of the fractal dimension of our simulated stars has yet to be conducted.  In any event, by keeping track explicitly of the G-actin monomers as the star forms, one enters the realm of diffusion limited aggregation (DLA)~\cite{witten}.  The angle-dependent branching and the bundling reorientation should lead to brand new cluster characteristics not yet observed in physical systems but, indeed, sought out by biological systems for reasons yet to be discovered. 

The authors would like to acknowledge helpful conversations with Andrea Liu, Ron Maimon, and Tatyana Svitkina during the early stages of this work. The authors gratefully acknowledge Louise Yang, an undergraduate summer intern who helped conduct some of the preliminary simulations in this work.  The authors would also like to acknowledge the hospitality of the Aspen Center for Physics where some of this work was completed. Finally, AG acknowledges support from the James S. McDonnell Foundation. 

\appendix
\section{Bundling due to bending}

To understand the role of bending, we construct the bundle size distribution for filaments emerging from an ordered two-dimensional array in which the filaments have the same initial angle of zero degrees with some regular spacing. Polymerizing filaments eventually bend in the transverse direction and ``find'' each other. It is the only way they can find each other.  We compute the collision probability (or rate) for a semiflexible polymer to first reach some transverse distance, i.e. to run into another growing semiflexible polymer, denoted by $\lambda^{\rm bend}_{i,k}$. As before, provided the angle is less then some critical angle, we assume realignment occurs.

To compute $\lambda^{bend}_{i,k}$, the Hamiltonian for a 
semiflexible polymer is given by
\be
\mathcal{H}=\frac{\kappa}{2}\int_{0}^{L}ds [\frac{\partial {\bf t}(s)}{\partial s}]^2,
\ee
where ${\bf t}(s)=\frac{\partial {\bf r}(s)}{\partial s}$ is the tangent vector 
of the curve ${\bf r}(s)$ and $\kappa$ is the bending rigidity, which is 
related to the persistence length, $L_p$, in two dimensions via $L_p=\frac{2\kappa}{k_b T}$~\cite{saito}.  Furthermore, the partition function is given by 
\be
Z(t_L,L|t_0,0)\propto \int_{{\bf t}(0)=t_0}^{{\bf t}(L)=t_L}\mathcal{D}[{\bf t}(s)]\delta(|{\bf t}(s)|-1)\exp(-\frac{\mathcal{H}}{k_B T}),
\ee
where the inextensibility condition is enforced by the delta function. Minimizing the semiflexible polymer paths yields an equation for $Z$, namely, 
\be
\frac{\partial Z}{\partial s}=\frac{1}{L_p}\frac{\partial^2 Z}{\partial \theta^2},
\label{eq:schrodinger}
\ee
where $\theta(s)$ is the angle between ${\bf t}(s)$ and a fixed reference axis. 
In order to obtain information about the position vector, ${\bf r}$, we   
replace the derivative on the left hand side of Eq. A3 with the convective derivative along the semiflexible polymer path, i.e. 
\be
(\frac{\partial}{\partial s}+\cos(\theta)\frac{\partial}{\partial x}+\sin(\theta)\frac{\partial}{\partial y})Z=\frac{1}{L_p}\frac{\partial^2 Z}{\partial \theta^2},
\label{eq:convective}
\ee
Assuming $\theta<<1$ and integrating over $x$, since we are only interested in the 
transverse fluctuations leading to bundling, simplifies Eq. A4 to 
\be
(\frac{\partial}{\partial s}+\theta\frac{\partial}{\partial y})Z=\frac{1}{L_p}\frac{\partial^2 Z}{\partial \theta^2}.
\label{eq:blah}
\ee

Defining $s'=s/L_p$ and $y'=y/L_p$ yields a Fokker-Planck equation
describing a particle undergoing random acceleration. In the absence of any
boundary condition, this equation is easily solved.  However, to compute a collision time one needs to 
consider this randomly accelerating particle in a finite interval $0<y<a$
where there are absorbing boundary conditions at either end of the interval.  To date, only the steady state absorption/collsion probability has been calculated by Bicout and Burkhardt~\cite{bicout}. If we assume that $\theta_0=0$ and the filament to be bundled starts halfway in between two other growing filaments, then the steady state absorption probability is 1/2. Converting this absorption probability to a collision rates yields $\lambda^{\rm bend}_{1,1}=\frac{1}{2}\frac{v_p \rho_1(0)}{L_p}$, where $\rho_1(0)$ is the initial spacing between filaments and we have assumed a constant concentration of G-actin.  Since we have averaged over $x$ and retain only the $y$-coordinate, we are considering a one-dimensional concentration.  To generalize 
$\lambda^{\rm bend}_{1,1}$ to bundles of size $i$ and $k$, $L_p(i)\propto i$ for a loosely packed bundle such that 
\begin{equation}
\lambda_{i,k}^{\rm bend}=\frac{\theta_c}{180^{\circ}}\frac{1}{2}\frac{v_p \rho_1(0)}{L_p}(i^{-1}+k^{-1}).
\end{equation}
This kernel is different from the case we analyzed in Section III because $\lambda^{\rm bend}_{i,k}$ decreases with increasing $i$ and $k$. The bundle persistance length increases as the bundle size increases making transverse fluctuations more difficult.

For the above kernel, the existence of self-similar solutions for the kinetic aggregation equation, in accordance with the dynamical scaling hypothesis, has recently been proved~\cite{fournier}.  In particular, 
\be
\rho_i(t)\sim \frac{1}{t}\tilde{\Psi}(\frac{i}{\sqrt{t}}).
\ee
Furthermore, $\tilde{\Psi}(x)$ is expected to scale as $x^{-\alpha}\exp(C x^{-\beta})$ for small $x$ such that $\rho_i(t)$ is nonmonotonic in $i$, unlike kernel studied in Section III.  We can test this result using the finite-dimensional simulations. 

In Sec. III we assumed previously that the cross-linkers are able to reorient, or bend, even the largest of bundles.  However, there is presumably a cross-linker concentration-dependent bundle size beyond which the cross-linkers cannot reorient the bundle in spite of the small angle collision.  As opposed to implementing a threshold bundle size for reorientation, one could implement a ``soft'' cutoff with $\lambda_{i,k}\sim (\sqrt{i}+\sqrt{k})(i^{-1}+k^{-1})$.  By incorporating this natural cutoff into the model, we can obtain a mean field bundle size distribution with a peak and an exponential tail.  We do not have to propose a rate for filaments leaving the bundle, which is unlikely given that the filaments are extended objects, as was done in Ref.~\cite{haviv2}, to obtain a peak.

\begin{figure}
\begin{center}
\epsfxsize=8cm
\epsfysize=6cm
\epsfbox{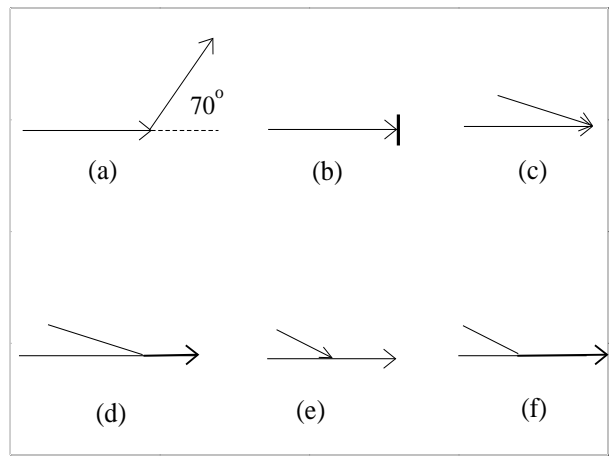}
\caption{Schematic diagrams of (a) branching at the tip with a 70 angle from the axis of the mother filament (b) capping (c) a tip-tip collsion and (d) the subsequent reorientation and bundling (e) a tip-side collision and (f) the subsequent reorientation.}
\label{fig:drawing}
\end{center}
\end{figure}

\begin{figure}
\begin{center}
\epsfxsize=7cm
\epsfysize=5cm 
\epsfbox{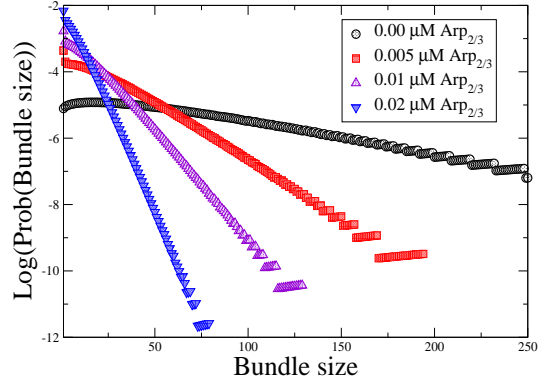}
\caption{Log-linear plot of final bundle size distributions for different concentrations of Arp2/3 with $\theta_c=20^{\circ}$.}
\label{fig:mf.bundlesizedist.arp.loglin}
\end{center}
\end{figure}

\begin{figure}
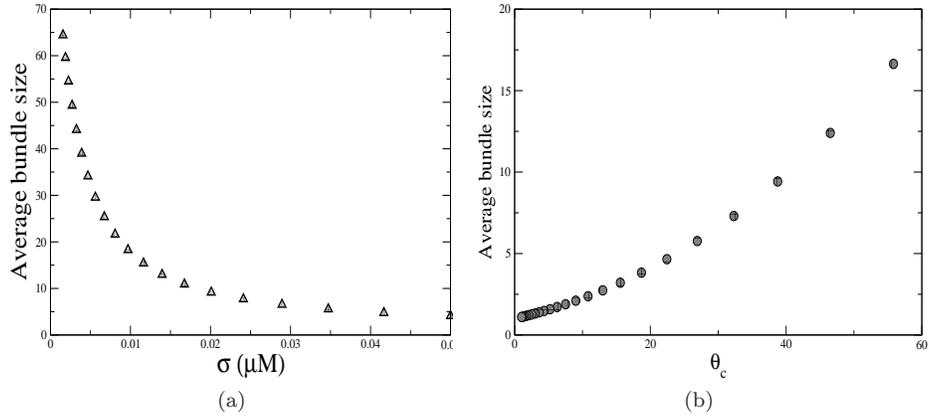

\begin{center}
\subfigure[]{\label{fig:mf.averagebundlesize.vs.arp-a}
\epsfxsize=6cm
\epsfysize=5cm 
\epsfbox{aggregate.averagebundlesize.versus.arp.reallyfinal.eps}}
\subfigure[]{\label{fig:mf.averagebundlesize.vs.thetac-b}
\epsfxsize=6cm
\epsfysize=5cm 
\epsfbox{aggregate.averagebundlesize.versus.thetac.reallyfinal.eps}}
\caption{(a) The average bundle size as a function of the concentration of Arp2/3 with $\theta_c=20^{\circ}$. (b) The average bundle size as a function of $\theta_c$ with $\sigma = 5\, nM$. }
\label{fig:mf.averagebundlesize.vs.arp}
\end{center}
\end{figure}

\begin{figure}
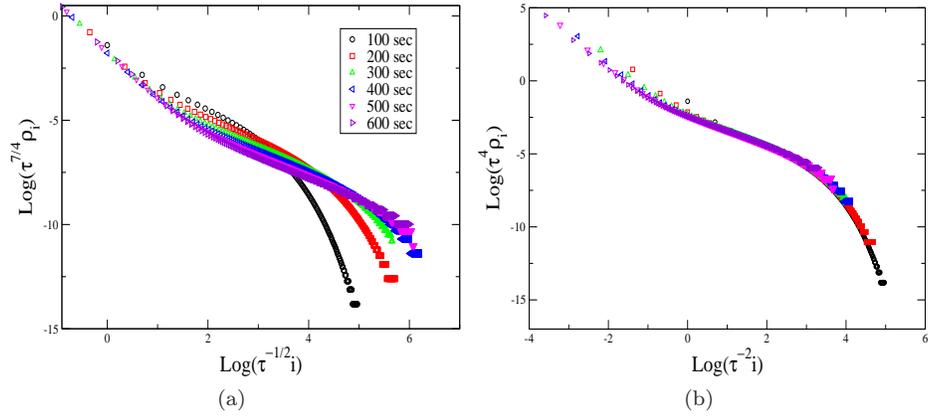

\begin{center}
\subfigure[]{\label{fig:mf.scalingcollapse.constantrhom-a}
\epsfxsize=6cm
\epsfysize=5cm 
\epsfbox{aggregate.reallyfinal.constantrhom.collapse1.eps}}
\subfigure[]{\label{fig:mf.scalingcollapse.constantrhom-b}
\epsfxsize=6cm
\epsfysize=5cm 
\epsfbox{aggregate.reallyfinal.constantrhom.collapse2.eps}}
\caption{(a) Scaling collapse for the intermediate bundle sizes with $\sigma=10\,nM$, $\theta_c=20^{\circ}$, and $\rho_m=10\, \mu M$.  The curves clearly illustrate convergence towards the proposed scaling in this region. (b) Scaling collapse for large bundle sizes with $\sigma=10\,nM$, $\theta_c=20^{\circ}$, and $\rho_m=10\,\mu M$. }
\label{fig:mf.scalingcollapse.constantrhom}
\end{center}
\end{figure}

\begin{figure}
\begin{center}
\subfigure[]{\label{fig:snapshot-a}
\epsfxsize=4cm
\epsfysize=4cm
\epsfbox{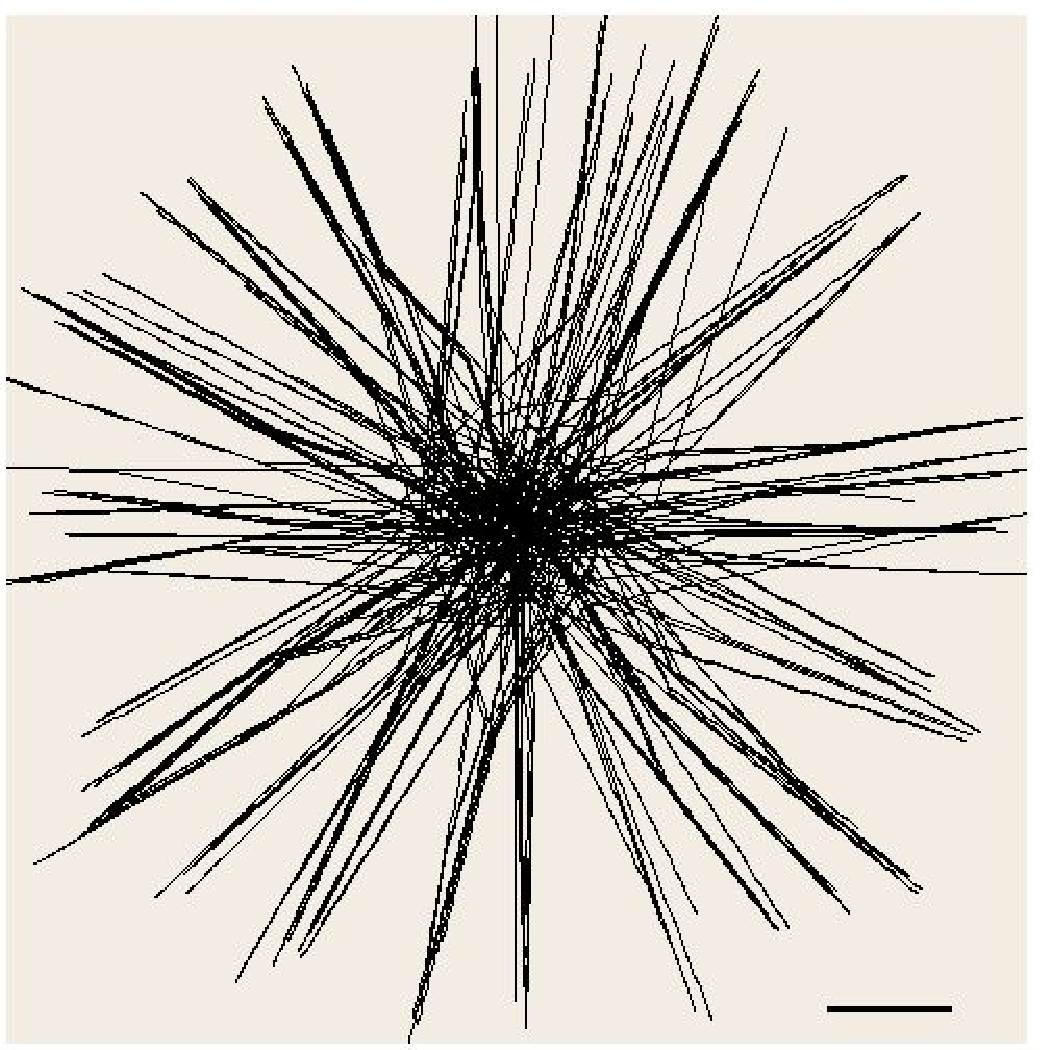}}
\subfigure[]{\label{fig:snapshot-b}
\epsfxsize=4cm
\epsfysize=4cm
\epsfbox{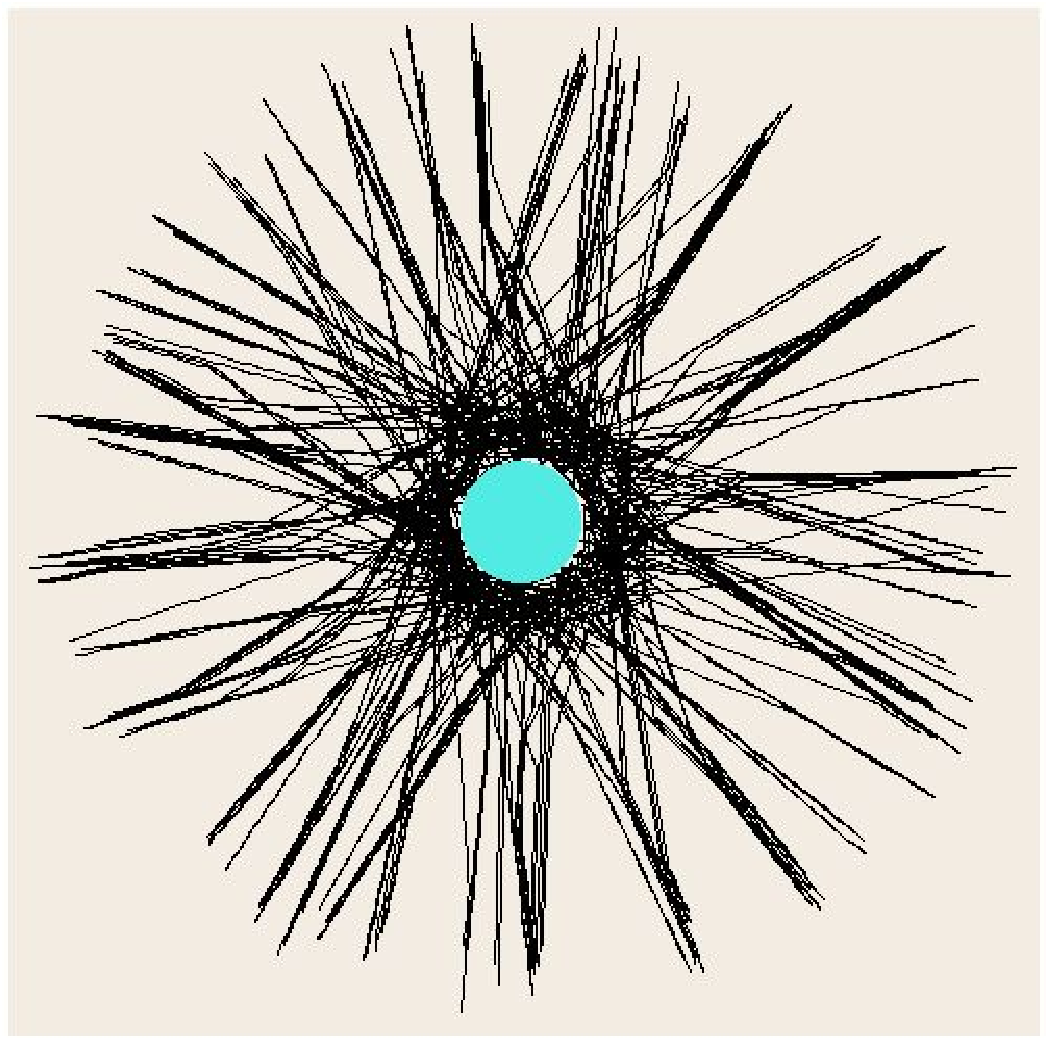}}
\caption{(a) Snapshot of the simulation with $R_b=0.0\, \mu m$. The line in the lower righthand corner is 1 micron. The remaining parameters are: $N_F=500$, $N_A=250$, $\theta_c=30^{\circ}$, and $R_i=0.6$, $d_G=0.003\,\mu m$, $d_X=0.017\,\mu m$, and $d_A=0.01\,\mu m$.   (b) Snapshot of the simulation with $R_b=0.5\,\mu m$ with the bead depicted in aqua (in color). The remaining parameters are the same as in (a) with the exception, $R_i=0.4\,\mu m$. }
\label{fig:snapshot}
\end{center}
\end{figure}

\begin{figure}
\begin{center}
\epsfxsize=7cm
\epsfysize=5cm
\epsfbox{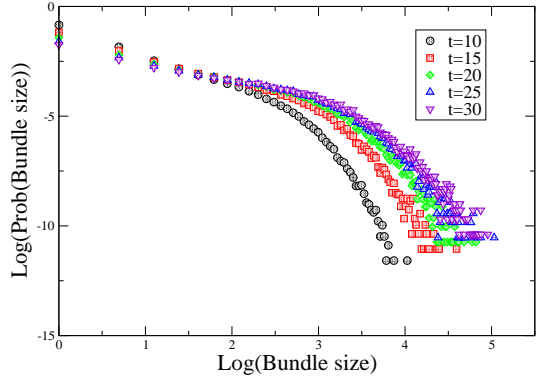}
\caption{Log-log plot the bundle size distribution at different times measured in simulation time steps. We use $N_F=500$, $N_A=250$, $R_b=0$, $R_i=0.6 \,\mu m$, $d_A=0.01\,\mu m$, $d_X=0.017\,\mu m$, $R_{cut}=0.75 \,\mu m$, $\theta_c=30^{\circ}$, and $d_f=0.03\, \mu m$.}
\label{fig:2d.star.nobead.bundlesizedist.dynamic}
\end{center}
\end{figure}

\begin{figure}
\begin{center}
\subfigure[]{\label{fig:2d.star.nobead.bundle1-a}
\epsfxsize=6cm
\epsfysize=5cm
\epsfbox{filo.star.nobead.bundle1.eps}}
\subfigure[]{\label{fig:2d.star.nobead.bundle1-b}
\epsfxsize=6cm
\epsfysize=5cm
\epsfbox{filo.star.nobead.bundle1.loglog.eps}}
\caption{(a) Log-linear plot of the bundle size distribution for different $\theta_c$s with $N_F=500$, $N_A=250$, $R_b=0$, $R_i=0.6 \,\mu m$, $d_A=0.01\,\mu m$, $d_X=0.017\,\mu m$, $R_{cut}=0.75 \,\mu m$, and $d_f=0.03\, \mu m$. (b) Log-log plot of Fig.~\ref{fig:2d.star.nobead.bundle1-a}. The line has a slope of -0.5.}
\label{fig:2d.star.nobead.bundle1}
\end{center}
\end{figure}

\begin{figure}
\begin{center}
\epsfxsize=7cm
\epsfysize=5cm
\epsfbox{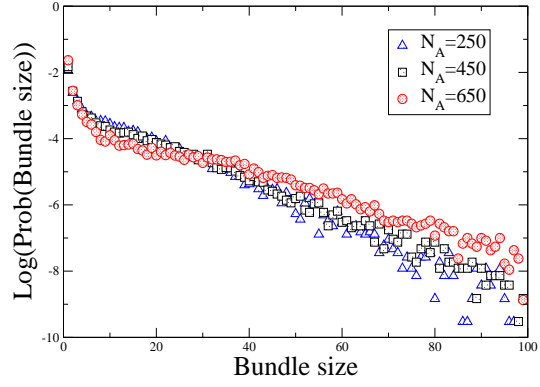}
\caption{Log-linear plot of the bundle size distribution for different $N_A$s with $N_F+N_A=750$, $R_b=0$, $R_i=0.6 \,\mu m$, $d_A=0.01\,\mu m$, $d_X=0.017\,\mu m$, $R_{cut}=0.75 \,\mu m$, $\theta_c=30^{\circ}$ and $d_f=0.03\, \mu m$.}
\label{fig:2d.star.nobead.bundle.arp}
\end{center}
\end{figure}

\begin{figure}
\begin{center}
\epsfxsize=7cm
\epsfysize=5cm
\epsfbox{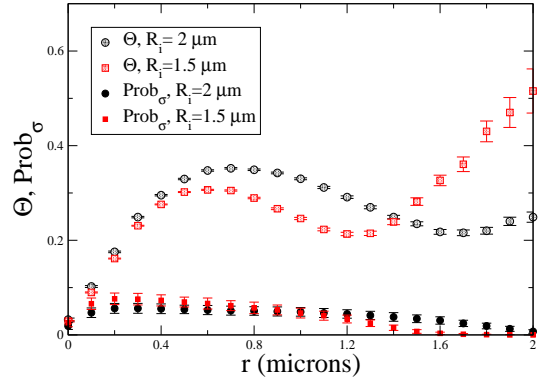}
\caption{$\Theta$ and the spatial distribution of the branched Arp2/3, $Prob_\sigma$, as a function of the radius for two different sets of parameters.  One data set is with  $N_F=100$, $N_A=650$, $R_i=2\, \mu m$ and $\theta_c=0$.  The second data set is with $N_F=100$, $N_A=650$, $R_i=1.5\, \mu m$, and $\theta_c=0$.}
\label{fig:2d.star.flux.calibrate}
\end{center}
\end{figure}

\begin{figure}
\begin{center}
\epsfxsize=7cm
\epsfysize=5cm
\epsfbox{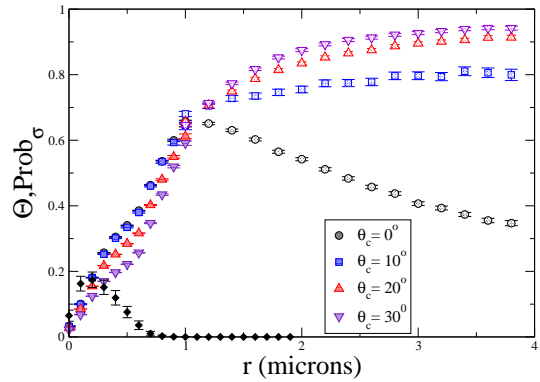}
\caption{Plot of $\Theta(r)$ for the parameters plotted in Fig.~\ref{fig:2d.star.nobead.bundle1} and the spatial distribution of branch points for $\theta_c=0^{\circ}$, which is plotted as filled diamonds.  }
\label{fig:2d.star.nobead.flux}
\end{center}
\end{figure}

\begin{figure}
\begin{center}
\epsfxsize=7cm
\epsfysize=5cm
\epsfbox{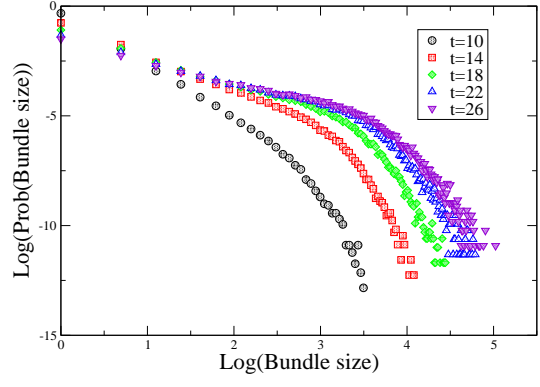}
\caption{Log-log plot the bundle size distribution at different simulation time steps. We use $N_F=500$, $N_A=250$, $R_b=0.5\,\mu m$, $R_i=0.4 \,\mu m$, $d_A=0.01\,\mu m$, $d_X=0.017\,\mu m$, $R_{cut}=0.75 \,\mu m$, $\theta_c=30^{\circ}$ and $d_f=0.03\, \mu m$.}
\label{fig:2d.star.bead.bundlesizedist.dynamic}
\end{center}
\end{figure}

\begin{figure}
\begin{center}
\subfigure[]{\label{fig:2d.star.bead.bundle1-a}
\epsfxsize=6cm
\epsfysize=5cm
\epsfbox{filo.star.bead.bundle1.eps}}
\subfigure[]{\label{fig:2d.star.bead.bundle1-b}
\epsfxsize=6cm
\epsfysize=5cm
\epsfbox{filo.star.bead.bundle1.loglog.eps}}
\caption{(a)Log-linear plot of the bundle size distribution for different $\theta_c$s with $N_F=500$, $N_A=250$, $R_b=0.5\,\mu m$, $R_i=1.2 \,\mu m$, $d_A=0.01\,\mu m$, $d_X=0.017\,\mu m$, $R_{cut}=0.75 \,\mu m$, and $d_f=0.03\, \mu m$. (b) Log-log plot of Fig.~\ref{fig:2d.star.bead.bundle1-a}. The line has a slope of -1.0.}
\label{fig:2d.star.bead.bundle1}
\end{center}
\end{figure}

\begin{figure}
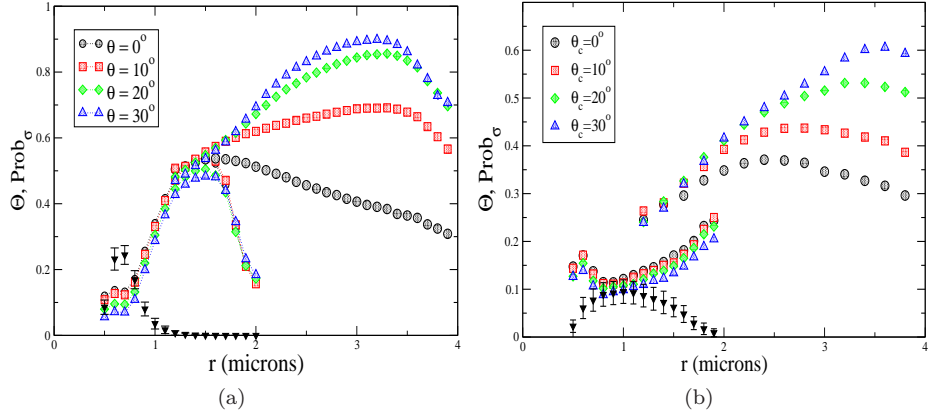

\begin{center}
\subfigure[]{\label{fig:2d.star.bead.flux-a}
\epsfxsize=6cm
\epsfysize=5cm
\epsfbox{filo.star.bead.fluxa.eps}}
\subfigure[]{\label{fig:2d.star.bead.flux-b}
\epsfxsize=6cm
\epsfysize=5cm
\epsfbox{filo.star.bead.flux.smallerdensity.eps}}
\caption{(a) $\Theta(r)$ for different $\theta_c$s with $R_b=0.5\,\mu m$ and $R_i=0.4\,\mu m$. We also plot $Prob_\sigma$ for the $\theta_c=0^{\circ}$ case, depicted in filled upside down triangles. (b)  $\Theta(r)$ for different $\theta_c$s with $R_b=0.5\,\mu m$ and $R_i=1.2\,\mu m$ and $Prob_\sigma$ for the $\theta_c=0^{\circ}$ case, again, depicted in filled upside down triangles. }
\label{fig:2d.star.bead.flux}
\end{center}
\end{figure}

\begin{figure}
\begin{center}
\epsfxsize=4cm
\epsfysize=4cm
\epsfbox{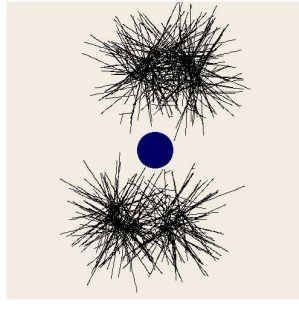}
\caption{Snapshot of the simulation with $R_b=0.5\, \mu m$, $R_i=0.4\,\mu m$, $\theta_c=30^{\circ}$, and $c=0.66 \,s^{-1}$.}
\label{fig:snapshot3}
\end{center}
\end{figure}

\begin{figure}
\begin{center}
\epsfxsize=7cm
\epsfysize=5cm
\epsfbox{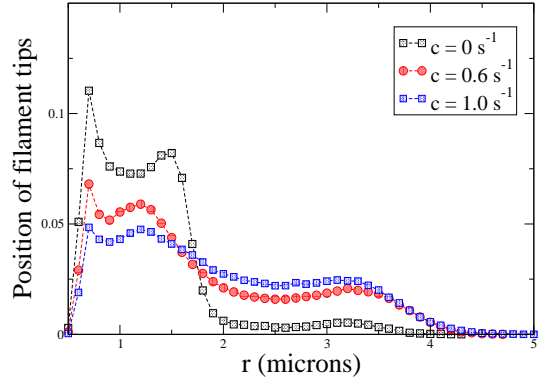}
\caption{Position of the filament tips as a function of the radius of the system for different capping rates.  Here, $R_i=0.3\,\mu m$ to accentuate the ratcheting effect.  }
\label{fig:2d.star.bead.capping}
\end{center}
\end{figure}

\begin{figure}
\begin{center}
\epsfxsize=7cm
\epsfysize=5cm 
\epsfbox{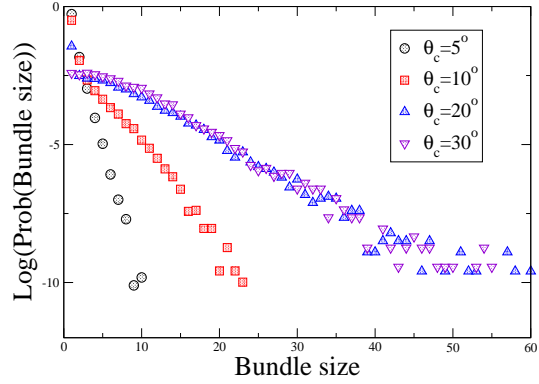}
\caption{Log-linear plot of the bundle size distribution for the ordered initial conditions. Here, $N_F=100$, the filament seeds start 0.1 microns apart with the same initial angle, and $d_X=0.017\,\mu m$. }
\label{fig:2d.ordered}
\end{center}
\end{figure}

\end{document}